\documentclass[12pt,a4paper]{article}

\usepackage[utf8]{inputenc}
\usepackage[T1]{fontenc}
\usepackage{lmodern}
\usepackage{amsmath,amssymb,amsfonts,amsthm,mathtools}
\usepackage{bm}
\usepackage{microtype}
\usepackage{enumitem}
\usepackage{geometry}
\usepackage{hyperref}

\usepackage[
  backend=biber,
  bibencoding=utf8,     
  style=numeric-comp,  
  sorting=none,        
  sortcites=true,      
  maxbibnames=99
]{biblatex}
\renewbibmacro{in:}{}

\setlist{nosep}

\hypersetup{
  colorlinks=true,
  linkcolor=black,
  citecolor=black,
  urlcolor=black,
  pdfauthor={Carlos Heredia},
  pdftitle={Nonlocal total derivatives and Euler-Lagrange equations}
}

\makeatletter
\renewcommand{\maketitle}{%
  \begin{center}
    {\Large\bfseries \@title\par}
    \vspace{0.9em}
    {\normalsize \@author\par}
    \vspace{0.6em}
    {\small \@date\par}
  \end{center}
  \vspace{0.8em}
}
\makeatother

\providecommand{\keywords}[1]{\par\vspace{0.5em}\noindent\textbf{Keywords:} #1\par\vspace{0.5em}}
\providecommand{\ack}[1]{\section*{Acknowledgements}#1}

\providecommand{\data}[1]{\section*{Data availability statement}#1}

\providecommand{\orcid}[1]{}

\theoremstyle{plain}
\newtheorem{theorem}{Theorem}[section]
\newtheorem{proposition}[theorem]{Proposition}
\newtheorem{corollary}[theorem]{Corollary}

\theoremstyle{definition}
\newtheorem{example}[theorem]{Example}
\theoremstyle{remark}

\newcommand{\R}{\mathbb{R}}

\newcommand{\dd}{\mathrm{d}}
\newcommand{\EL}{\mathcal{E}}
\newcommand{\Lag}{\mathcal{L}}

\newcommand{\D}{D}

\newcommand{\conv}{\ast}
\newcommand{\erf}{\operatorname{erf}}

\title{When Nonlocal Total Derivatives \\ Are Not Variationally Trivial}
\author{Carlos Heredia\\
\small IAMM Research, Department of Applied Artificial Intelligence, DAMM,\\
\small Carrer del Rossello 515, 08025 Barcelona, Catalonia, Spain\\
\small \texttt{carlosherediapimienta@gmail.com}}
\date{}

\begin{document}

\maketitle

\begin{abstract}
We study total time derivatives in infinite-order and nonlocal Lagrangian mechanics. In the formal infinite-order setting, such terms remain variationally trivial. For nonlocal functionals on trajectories, however, exactness in the form \(\Lag=\D W\) is not sufficient to ensure a vanishing nonlocal Euler-Lagrange expression. We exhibit a smooth convolution-type counterexample.
\end{abstract}


\section{Introduction}

Nonlocal Lagrangians and infinite-derivative actions appear in several gravitational and high-energy contexts, including nonlocal gravity, string-inspired effective actions, dispersive media and higher-derivative oscillator models \cite{Tetrode1922,Fokker,WheelerFeynman,Yukawa,Moller,JLlosaMT,Witten1986,EliezerWoodard1989,Biswas2012,HerediaKolar2022,BoosKolar2021,CalcagniModesto2014,CapozzielloBajardi2022,Barvinsky2012,MaggioreMancarella2014,ModestoRachwal2014,Tomboulis2015,BasCalcagniRachwal2024,KoshelevKumarStarobinsky2023,Mashhoon2011}. A recurring technical point in these theories is the status of boundary terms. In a local finite-order variational problem, a total time derivative changes the action by an endpoint contribution and leaves the Euler-Lagrange (EL) equations unchanged \cite{Landau}. The same conclusion is often used in formal infinite-derivative manipulations \cite{Biswas_2013, Erbin_2022}.

For nonlocal Lagrangians, this inference is not automatic. In the functional formulation of nonlocal mechanics, it was already observed that a Lagrangian may be written as the total derivative of a nonlocal functional without having identically vanishing Euler-Lagrange equations \cite{HerediaThesis2023}. Related results for nonlocal field theories, where total divergences require additional asymptotic conditions in order to be variationally trivial, were developed in \cite{Heredia2022, Heredia_2024_U1}. The underlying reason is that a nonlocal functional may depend on an extended portion of the trajectory, or on the full trajectory. Consequently, a boundary contribution need not depend only on endpoint data, but may retain a nontrivial dependence on variations of the trajectory.

This motivates the structural question considered in this work:
\begin{quote}
Under what conditions does \(L=DW\) imply variational triviality in formal infinite-order calculus and in nonlocal functional mechanics?
\end{quote}

To address this question, we first derive algebraic divisibility criteria in the formal theory.  We then revisit the nonlocal boundary criterion \cite{Heredia2022,HerediaThesis2023,Heredia_2024_U1} and the convolutional construction \cite{Heredia2021a,Heredia2026NM}, specializing
the latter to a shifted Gaussian kernel and an explicit trajectory for which the nonlocal Euler-Lagrange expression does not vanish. We finally compare this functional result with its formal infinite-derivative counterpart.

Section~\ref{sec:formal} develops the formal criteria, while Section~\ref{sec:nonlocal} reviews the nonlocal variational framework and applies the convolutional model to the shifted Gaussian example. Appendix~\ref{app:formal-convolution} analyses the corresponding formal infinite-derivative representation, and Section~\ref{sec:discussion} summarizes the distinction between formal and functional variational triviality.

\section{Formal infinite-order total derivatives}
\label{sec:formal}
Let $q_k:=D^kq$, with $k\geq0$, denote the formal jet variables, so that
\[
Dq_k=q_{k+1}.
\]
A finite-order differential polynomial depends on only finitely many of these variables, say \(q_0,\ldots,q_M\). Here we enlarge this setting in order to include formal expressions containing derivatives of arbitrarily high order. For \(r,N\geq0\), let
\[
\mathcal A_{r,N}
:=
\operatorname{span}
\left\{
q_{i_1}\cdots q_{i_r}
:\,
i_1+\cdots+i_r=N
\right\},
\]
with
\[
\mathcal A_{0,0}=\mathbb R,
\qquad
\mathcal A_{0,N}=\{0\}
\quad (N>0).
\]
Here \(r\) counts the number of jet factors in a monomial, while \(N\) is the total number of derivatives carried by them. For example, $q_1q_3\in\mathcal A_{2,4}$, or $q_0q_2^2\in\mathcal A_{3,4}$. For fixed \(r\) and \(N\), the space \(\mathcal A_{r,N}\) is finite-dimensional.

We define the formal jet algebra by
\[
\widehat{\mathcal A}
:=
\prod_{r,N\geq0}\mathcal A_{r,N},
\qquad
\widehat{\mathcal A}_r
:=
\prod_{N\geq0}\mathcal A_{r,N}.
\]
Thus an element of \(\widehat{\mathcal A}\) may contain infinitely many differential orders. For instance, $W^{(M)} = \frac12\sum_{n=0}^{M}q_n^2$ is a finite-order differential polynomial, whereas $W = \frac12\sum_{n=0}^{\infty}q_n^2 \in\widehat{\mathcal A}_2$ is a infinite-order formal element. The latter is understood purely formally.

The total derivative is well defined because
\[
D:\mathcal A_{r,N}\longrightarrow\mathcal A_{r,N+1}.
\]
For example, $D\left( \frac12\sum_{n=0}^{\infty}q_n^2 \right) = \sum_{n=0}^{\infty}q_nq_{n+1}$ as a formal identity.

The formal Euler-Lagrange operator is
\[
E(L)
:=
\sum_{k\geq0}
(-D)^k\frac{\partial L}{\partial q_k}.
\]
It is also well defined componentwise. Indeed, for
\(P\in\mathcal A_{r,N}\),
\[
\frac{\partial P}{\partial q_k}=0
\qquad\text{for }k>N,
\]
and
\[
D^k\frac{\partial P}{\partial q_k}
\in\mathcal A_{r-1,N}.
\]
Consequently, only the terms with \(0\leq k\leq N\) can contribute to \(E(P)\), and
\[
E:\mathcal A_{r,N}\longrightarrow\mathcal A_{r-1,N}.
\]

We now relate this setting to the symbolic description. The idea is to encode the jet indices as powers of auxiliary variables, so that increasing the order of a time derivative of \(q\) corresponds to multiplication by a formal variable. In particular, we associate
\[
q_{i_1}\cdots q_{i_r}
\quad\longleftrightarrow\quad
x_1^{i_1}\cdots x_r^{i_r},
\]
where \(x_1,\ldots,x_r\) are auxiliary formal variables. For a homogeneous expression
\[
L_r
=
\frac{1}{r!}
\sum_{i_1,\ldots,i_r\geq0}
C^{(r)}_{i_1\cdots i_r}
q_{i_1}\cdots q_{i_r},
\]
we therefore encode the coefficients \(C^{(r)}_{i_1\cdots i_r}\) in the symmetric generating series
\[
C_r(x_1,\ldots,x_r)
:=
\sum_{i_1,\ldots,i_r\geq0}
C^{(r)}_{i_1\cdots i_r}
x_1^{i_1}\cdots x_r^{i_r}\,.
\]
We shall refer to \(C_r(x_1,\ldots,x_r)\) as the symbol associated with \(L_r\). For a finite-order differential polynomial, \(C_r\) is a polynomial, whereas in the setting considered here it may be a formal power series.

The usefulness of this representation follows from \(Dq_i=q_{i+1}\): increasing the derivative order of the \(a\)-th jet factor by one corresponds to multiplying its symbolic representation by \(x_a\). Indeed, for finite-order differential polynomials, Olver and Shakiban showed that this gives
\[
D
\quad\longleftrightarrow\quad
x_1+\cdots+x_r,
\]
while the symbol of \(E(L_r)\) is obtained by evaluating \(C_r(x_1,\ldots,x_r)\) at
\[
x_1=-(x_2+\cdots+x_r)
\]
\cite[Lemma~5]{OlverShakiban1978}. We show below that the same symbolic relations remain valid in the formal infinite-derivative setting considered here, where derivatives of arbitrarily high order may occur. This provides the formal infinite-order framework that will later be compared with nonlocal functionals in Section~\ref{sec:nonlocal}.

\subsection{Quadratic formal Lagrangians}

We begin with the quadratic case.

\begin{theorem}[Quadratic infinite-order criterion]\label{th:qic}
Let
\[
L=
\frac12
\sum_{m,n\geq0}
C_{mn}q_mq_n
\in\widehat{\mathcal A}_2,
\qquad
C_{mn}=C_{nm},
\]
and let
\[
C(x,y)
:=
\sum_{m,n\geq0}
C_{mn}x^my^n
\in\mathbb R[[x,y]]
\]
be its symmetric symbol. Then the following conditions are equivalent:
\begin{align}
&\text{there exists }W\in\widehat{\mathcal A}_2
\text{ such that }L=DW, \label{eq:quad-a}\\
&C(x,y)\in(x+y), \label{eq:quad-b}\\
&C(x,-x)=0. \label{eq:quad-c}
\end{align}
Equivalently, for every \(N\geq0\),
\[
\sum_{m+n=N}(-1)^n C_{mn}=0.
\]
When these conditions hold, the symbol of a primitive \(W\) is
\[
W(x,y)=\frac{C(x,y)}{x+y}
\in\mathbb R[[x,y]].
\]
\end{theorem}

\begin{proof}
Let
\[
W=
\frac12
\sum_{i,j\geq0}
W_{ij}q_iq_j,
\qquad
W_{ij}=W_{ji},
\]
and let
\[
W(x,y)
=
\sum_{i,j\geq0}W_{ij}x^iy^j
\]
be its symbol. Since \(Dq_i=q_{i+1}\),
\[
DW
=
\frac12
\sum_{i,j\geq0}
W_{ij}
\left(
q_{i+1}q_j+q_iq_{j+1}
\right).
\]
In the symbol, increasing the first jet index by one corresponds to multiplication by \(x\), while increasing the second by one corresponds to multiplication by \(y\). Therefore the symbol of \(DW\) is $(x+y)W(x,y)$. Hence, if \(L=DW\), then their symbols satisfy
\[
C(x,y)=(x+y)W(x,y),
\]
so \(C(x,y)\) is divisible by \(x+y\). Conversely, suppose that
\[
C(x,y)=(x+y)B(x,y)
\]
for some \(B(x,y)\in\mathbb R[[x,y]]\). Let \(W\in\widehat{\mathcal A}_2\) be the quadratic formal element whose symbol is \(B(x,y)\). Then the symbol of \(DW\) is
\[
(x+y)B(x,y)=C(x,y).
\]
Since \(L\) and \(DW\) have the same symbol, they are equal. Therefore
\[
L=DW
\quad\Longleftrightarrow\quad
C(x,y)\in(x+y).
\]

All these identities are understood coefficientwise. In particular, for a fixed total degree \(N\), only the finitely many pairs \((m,n)\) satisfying \(m+n=N\) contribute. Thus no convergence assumption is involved.

It remains to relate divisibility by \(x+y\) to the condition \(C(x,-x)=0\). Introduce the formal variables $u=x+y$, and $v=y$, so that $x=u-v$. This defines an invertible linear change of formal variables. Hence
\[
C(u-v,v)
=
\sum_{j\geq0}u^j A_j(v)
\]
for uniquely determined \(A_j(v)\in\mathbb R[[v]]\), where
\[
A_j(v)
:=
\sum_{\substack{m\geq j\\ n\geq0}}
C_{mn}
\binom{m}{j}
(-1)^{m-j}
v^{m-j+n}.
\]
The substitution \(x=-y\) is equivalent to \(u=0\). Therefore
\[
C(x,-x)=0 \quad\Longleftrightarrow\quad A_0(v)=0.
\]
If \(A_0(v)\equiv0\), then
\[
C(u-v,v)
=
\sum_{j\geq1}u^jA_j(v)
=
u\sum_{j\geq1}u^{j-1}A_j(v).
\]
Hence \(C(u-v,v)\) is divisible by \(u\). Conversely, if \(C(u-v,v)=uB(u,v)\), then setting \(u=0\) gives \(A_0(v)=0\). Since \(u=x+y\), this is equivalent to divisibility of \(C(x,y)\) by \(x+y\).

Finally,
\[
C(x,-x)
=
\sum_{N\geq0}
\left(
\sum_{m+n=N}(-1)^n C_{mn}
\right)x^N,
\]
so \(C(x,-x)=0\) is equivalent to
\[
\sum_{m+n=N}(-1)^n C_{mn}=0
\]
for every \(N\geq0\).
\end{proof}

\begin{corollary}[Quadratic variational triviality]
Under the assumptions of the previous theorem,
\[
L=DW
\qquad\Longrightarrow\qquad
E(L)=0.
\]
Consequently, \(L_{\mathrm{ref}}\) and \(L_{\mathrm{ref}}+DW\) have the same formal Euler-Lagrange expression.
\end{corollary}

\begin{proof}
For
\[
L=
\frac12
\sum_{m,n\geq0}C_{mn}q_mq_n,
\]
symmetry of \(C_{mn}\) gives
\[
\frac{\partial L}{\partial q_k}
=
\sum_{n\geq0}C_{kn}q_n.
\]
Therefore
\[
E(L)
=
\sum_{k,n\geq0}
(-1)^k C_{kn}q_{k+n}.
\]
Collecting all terms of total derivative order \(N\),
\[
E(L)
=
\sum_{N\geq0}
\left(
\sum_{k=0}^{N}
(-1)^k C_{k,N-k}
\right)q_N.
\]
For each fixed \(N\), the inner sum is finite, so the expression is well defined in this algebra. By the Theorem \ref{th:qic},
\[
\sum_{m+n=N}(-1)^n C_{mn}=0.
\]
Writing \(n=N-k\),
\[
\sum_{m+n=N}(-1)^n C_{mn}
=
(-1)^N
\sum_{k=0}^{N}
(-1)^k C_{k,N-k}.
\]
Hence every coefficient of \(E(L)\) vanishes and therefore
\[
E(L)=0.
\]
The last statement follows from the linearity of \(E\).
\end{proof}

Two elementary examples clarify the meaning of this criterion. For \(L=q\dot q=q_0q_1\), one has \(C(x,y)=x+y\), so \(W=\frac12q^2\). By contrast, for \(L=\frac12\dot q^2\), one has \(C(x,y)=xy\) and \(C(x,-x)=-x^2\neq0\), so it is not a formal quadratic total derivative.

\subsection{Homogeneous formal Lagrangians}

We now consider homogeneous expressions of arbitrary polynomial degree. In the finite-order case treated in~\cite{OlverShakiban1978}, the symbol is a polynomial and only finitely many derivative orders are present. In the formal infinite-derivative setting considered here, \(C_r\) may instead be a formal power series involving arbitrarily high differential orders. Nevertheless, multiplication by $x_1+\cdots+x_r$ remains well defined degree by degree: the component of total degree \(N+1\) depends only on the component of degree \(N\). Hence the symbolic action of \(D\) extends without ambiguity to the infinite-derivative case.

\begin{theorem}[Homogeneous infinite-order criterion]\label{th:hic}
Let \(r\geq1\) and let
\[
L_r
=
\frac{1}{r!}
\sum_{i_1,\ldots,i_r\geq0}
C^{(r)}_{i_1\cdots i_r}
q_{i_1}\cdots q_{i_r}
\in\widehat{\mathcal A}_r,
\]
where \(C^{(r)}\) is symmetric in its indices. Let
\[
C_r(x_1,\ldots,x_r)
:=
\sum_{i_1,\ldots,i_r\geq0}
C^{(r)}_{i_1\cdots i_r}
x_1^{i_1}\cdots x_r^{i_r}
\]
be the associated symmetric formal power series. Then the following conditions are equivalent:
\begin{align}
&E(L_r)=0, \label{eq:hom-a}\\
&C_r(-x_2-\cdots-x_r,x_2,\ldots,x_r)=0, \label{eq:hom-b}\\
&C_r(x_1,\ldots,x_r)
\in
(x_1+\cdots+x_r), \label{eq:hom-c}\\
&\text{there exists }W_r\in\widehat{\mathcal A}_r
\text{ such that }L_r=DW_r. \label{eq:hom-d}
\end{align}
When these conditions hold, the symbol of \(W_r\) is
\[
B_r(x_1,\ldots,x_r)
=
\frac{C_r(x_1,\ldots,x_r)}
{x_1+\cdots+x_r}.
\]
\end{theorem}

\begin{proof}
We first determine the action of \(D\) on the symbol. For a monomial $q_{i_1}\cdots q_{i_r}$, the total derivative gives
\[
D(q_{i_1}\cdots q_{i_r})
=
\sum_{a=1}^{r}
q_{i_1}\cdots q_{i_a+1}\cdots q_{i_r}.
\]
At the symbolic level, increasing the index \(i_a\) by one corresponds to multiplication by \(x_a\). Hence
\[
D
\quad\longleftrightarrow\quad
x_1+\cdots+x_r.
\]
Therefore
\[
L_r=DW_r \quad  \Longleftrightarrow \quad 
C_r(x_1,\ldots,x_r)
=
(x_1+\cdots+x_r)
B_r(x_1,\ldots,x_r),
\]
where \(B_r\) is the symbol of \(W_r\). Thus the argument of the Theorem \ref{th:qic} extends directly, with \(x+y\) replaced by \(x_1+\cdots+x_r\), proving the equivalence between \eqref{eq:hom-c} and \eqref{eq:hom-d}.

We next consider the Euler operator. Symmetry of \(C^{(r)}\) gives
\[
\frac{\partial L_r}{\partial q_k}
=
\frac{1}{(r-1)!}
\sum_{i_2,\ldots,i_r\geq0}
C^{(r)}_{k i_2\cdots i_r}
q_{i_2}\cdots q_{i_r}.
\]
Set $S:=x_2+\cdots+x_r.$ Applying \(D^k\) to a homogeneous expression of degree \(r-1\) corresponds to multiplication of its symbol by \(S^k\). Therefore the symbol of $(-D)^k\partial L_r / \partial q_k$ is obtained by multiplying by \((-S)^k\). Summing over \(k\), the symbol of \(E(L_r)\) is
\[
\sum_{k\geq0}
(-S)^k
\sum_{i_2,\ldots,i_r\geq0}
C^{(r)}_{k i_2\cdots i_r}
x_2^{i_2}\cdots x_r^{i_r},
\]
which is precisely $C_r(-S,x_2,\ldots,x_r)$. Thus
\[
E(L_r)=0 \quad \Longleftrightarrow \quad 
C_r(-S,x_2,\ldots,x_r)=0,
\]
proving the equivalence between
\eqref{eq:hom-a} and \eqref{eq:hom-b}.

The substitution above is well defined for formal power series. Indeed, it preserves the total degree, so every coefficient of $C_r(-S,x_2,\ldots,x_r)$ depends on only finitely many coefficients of \(C_r\).

It remains to prove the equivalence between \eqref{eq:hom-b} and \eqref{eq:hom-c}. Introduce $u=x_1+x_2+\cdots+x_r$, then $x_1=u-S.$ This is an invertible linear change of formal variables, and therefore
\[
C_r(u-S,x_2,\ldots,x_r)
=
\sum_{j\geq0}
u^j A_j(x_2,\ldots,x_r)
\]
where 
\[
A_j(x_2,\ldots,x_r)
=
\sum_{\substack{i_1\geq j\\ i_2,\ldots,i_r\geq0}}
C^{(r)}_{i_1\cdots i_r}
\binom{i_1}{j}
(-1)^{i_1-j}
S^{\,i_1-j}
x_2^{i_2}\cdots x_r^{i_r}\,.
\]
Thus \(A_j\in\mathbb R[[x_2,\ldots,x_r]]\). Since the substitution \(x_1=-S\) is equivalent to setting \(u=0\), we have
\[
C_r(-S,x_2,\ldots,x_r)
=
A_0(x_2,\ldots,x_r).
\]
Therefore,
\[
C_r(-S,x_2,\ldots,x_r)=0
\quad\Longleftrightarrow\quad
A_0(x_2,\ldots,x_r)=0.
\]
If \(A_0=0\), the expansion starts at first order in \(u\),
\[
C_r(u-S,x_2,\ldots,x_r)
=
\sum_{j\geq1}u^jA_j(x_2,\ldots,x_r)
=
u\sum_{j\geq1}u^{j-1}A_j(x_2,\ldots,x_r),
\]
and is therefore divisible by \(u\). Conversely, divisibility by \(u\) forces the constant term \(A_0\) to vanish. Since $u=x_1+\cdots+x_r,$ we conclude that
\[
C_r(-S,x_2,\ldots,x_r)=0
\quad\Longleftrightarrow\quad
C_r(x_1,\ldots,x_r)\in(x_1+\cdots+x_r).
\]
This completes the proof.
\end{proof}

\subsection{Mixed formal Lagrangians}

We finally allow different polynomial degrees to occur simultaneously.

\begin{theorem}[Completed formal criterion]\label{th:cfc}
Let
\[
L=\sum_{r\geq0}L_r
\in\widehat{\mathcal A},
\qquad
L_r\in\widehat{\mathcal A}_r,
\]
and let \(C_r(x_1,\ldots,x_r)\) denote the symmetric symbol of \(L_r\) for \(r\geq1\). Then
\[
E(L)=0
\quad\Longleftrightarrow\quad
E(L_r)=0
\qquad
\text{for every }r\geq1.
\]
Equivalently,
\[
E(L)=0
\quad\Longleftrightarrow\quad
C_r(x_1,\ldots,x_r)
\in
(x_1+\cdots+x_r)
\qquad
\text{for every }r\geq1.
\]
When these conditions hold, there exists
\[
W=\sum_{r\geq1}W_r
\in\widehat{\mathcal A}
\]
such that
\[
L=L_0+DW.
\]
In the autonomous formal algebra considered here,
\[
L_0\in\mathbb R.
\]
\end{theorem}

\begin{proof}
Let us write more explicitly $L=\sum_{r,N\geq0}L_{r,N}$ with $L_{r,N}\in\mathcal A_{r,N}$. By construction, $E: \mathcal A_{r,N} \longrightarrow \mathcal A_{r-1,N}$. Hence the component of \(E(L)\) belonging to \(\mathcal A_{s,N}\) can receive a contribution only from \(L_{s+1,N}\). Therefore the terms \(E(L_r)\) belong to different polynomial degrees and cannot cancel each other. It follows that
\[
E(L)=0 \Longleftrightarrow
E(L_r)=0
\]
for every \(r\geq1\).

Applying the Theorem \ref{th:hic} to each homogeneous component \(L_r\), we have
\[
E(L_r)=0
\quad\Longleftrightarrow\quad
C_r(x_1,\ldots,x_r)
\in
(x_1+\cdots+x_r).
\]
This proves the second equivalence.

Whenever these conditions hold, Theorem~\ref{th:hic} gives, for every \(r\geq1\), an element
\[
W_r\in\widehat{\mathcal A}_r
\qquad\text{such that}\qquad
L_r=DW_r.
\]
Define $W:=\sum_{r\geq1}W_r$. By the definition of \(\widehat{\mathcal A}\), this formal sum defines a well-defined element of \(\widehat{\mathcal A}\). Moreover, since \(D\) acts linearly and componentwise and satisfies $D:\mathcal A_{r,N}\longrightarrow\mathcal A_{r,N+1}$, it can be applied term by term to the formal sum. Hence
\[
DW
=
\sum_{r\geq1}DW_r
=
\sum_{r\geq1}L_r.
\]
Therefore,
\[
L=L_0+DW.
\]

Finally, the degree-zero component of the autonomous formal algebra contains no jet variables and no explicit time dependence. Hence
\[
L_0\in\mathbb R.
\]
\end{proof}

This provides a useful interpretation of standard manipulations in infinite-derivative theories, where integrations by parts and total derivatives are often performed at the level of formal derivative expansions; see, for example, the time-dependent reduction in \cite{Erbin_2022}. Theorem~\ref{th:cfc} shows that, within the formal infinite-derivative framework considered here, a term of the form \(DW\), with \(W\in\widehat{\mathcal A}\), is variationally trivial: its Euler-Lagrange expression vanishes degree by degree.

This conclusion is purely formal. The primitive $W=\sum_{r\geq1}W_r$ may contain derivatives of arbitrarily high order, and the identity $E(DW)\equiv0$ is understood coefficientwise in the formal jet algebra. No convergence of the corresponding series on an actual trajectory is required. If, instead, an infinite formal expression defines a functional on a space of trajectories, the situation is more subtle. Varying the formal series term by term does not necessarily give the variation of the resulting functional. Thus, whether variational triviality is preserved in this setting requires further analysis, which we provide in the next section.

\section{Nonlocal Euler-Lagrange operator}\label{sec:nonlocal}

Following the trajectory-space formulation of nonlocal mechanics \cite{Heredia2021a, HerediaThesis2023, Heredia2026NM}, we take the kinematic space $\mathcal K$ to be a suitable space of admissible trajectories, for instance $C^k(\mathbb R,\mathbb R^m)$ or an appropriate subspace thereof, depending on the Lagrangian under consideration. We assume that $\mathcal K$ is invariant under time translations and admits smooth, compactly supported variations. Moreover, we assume that the required functional derivatives exist and admit kernel representations, and that the limits and interchanges of variation and integration used below are justified.
Additional admissibility conditions for the convolutional example will be specified separately.

For $q\in\mathcal K$, define the translated trajectory
\[
T_tq(\tau):=q(t+\tau).
\]
A nonlocal Lagrangian, allowing explicit time dependence, is a functional assigning the value $\Lag(T_tq,t)$. For fixed \(t\), it may depend on the values of \(q\) at times other than \(t\), possibly over the entire real line, rather than only on the finite jet of \(q\) at \(t\).

Following the truncated-action approach of \cite{HerediaThesis2023, Heredia2026NM}, we consider the symmetric family actions
\[
S_R[q]:=
\int_{-R}^{R}\dd t\,\Lag(T_tq,t),
\qquad R>0.
\]
For a smooth compactly supported variation $\delta q\in C_c^\infty(\mathbb R,\mathbb R^m)$, we define
\[
\delta S_R[q;\delta q]
:=
\left.
\frac{\dd}{\dd\varepsilon}
S_R[q+\varepsilon \delta q]
\right|_{\varepsilon=0}.
\]
We then consider
\[
\delta S[q;\delta q]
:=
\lim_{R\to\infty}\delta S_R[q;\delta q],
\]
provided the limit exists, and impose the stationarity condition
\[
\delta S[q;\delta q]=0
\]
for all admissible compactly supported variations. Writing
\[
\lambda(q,t,\sigma):=
\frac{\delta\Lag(T_tq,t)}{\delta q(\sigma)},
\]
the first variation can be expressed, under the assumptions stated above, as
\[
\delta S[q;\delta q]
=
\int_{\mathbb R}\dd\sigma\,
\Psi(q,\sigma)\delta q(\sigma),
\qquad
\Psi(q,\sigma):=
\lim_{R\to\infty}
\int_{-R}^{R}\dd t\,\lambda(q,t,\sigma),
\]
where $\Psi$ denotes the nonlocal Euler-Lagrange operator. Therefore, the dynamical space is
\[
\mathcal D
:=
\left\{
q\in\mathcal K:\Psi(q,\cdot)=0
\right\}
\subset\mathcal K.
\]

A nonlocal total derivative is of the form
\begin{equation}\label{eq:nonlocal-total-derivative}
\Lag(T_tq,t)=\D W(T_tq,t),
\end{equation}
where
\begin{equation}\label{eq:derivativeW}
 \D W(T_tq,t):=
\left.
\frac{\dd}{\dd\varepsilon}
W(T_{t+\varepsilon}q,t+\varepsilon)
\right|_{\varepsilon=0}.   
\end{equation}

The following observation was already established in the nonlocal Lagrangian formalism of \cite{HerediaThesis2023}. We recall it here because it makes clear why the mere existence of a nonlocal primitive cannot, by itself, characterize variational triviality.

\begin{theorem}[Universal nonlocal primitive]\label{thm:universal-nonlocal-primitive}
Let $\mathcal K$ be invariant under time translations, and assume that, for every $q\in\mathcal K$, the function $t\mapsto\Lag(T_tq,t)$ is continuous. Then there exists a nonlocal functional \(W\) such that
\[
\Lag(T_tq,t)=\D W(T_tq,t).
\]
One possible choice is
\begin{equation}\label{eq:universal-W}
W(T_tq,t)=\int_0^t \dd s\,\Lag(T_sq,s).
\end{equation}
\end{theorem}

\begin{proof}
Using \eqref{eq:derivativeW} and \eqref{eq:universal-W}, we have
\[
W(T_{t+\varepsilon}q,t+\varepsilon)
=
\int_0^{t+\varepsilon}\dd s\,\Lag(T_sq,s).
\]
Therefore
\[
\D W(T_tq,t)
=
\left.
\frac{\dd}{\dd\varepsilon}
\int_0^{t+\varepsilon}\dd s\,\Lag(T_sq,s)
\right|_{\varepsilon=0}
=
\Lag(T_tq,t).
\]
\end{proof}
Thus, in the nonlocal case, the relevant question is not merely whether a Lagrangian can be written as a total derivative---since any sufficiently regular Lagrangian admits a representation as a nonlocal total derivative---but whether the associated boundary contribution is variationally trivial. Indeed, integrating \eqref{eq:nonlocal-total-derivative} over a finite interval
\([-R,R]\) gives
\[
S_R[q]=W(T_Rq,R)-W(T_{-R}q,-R),
\]
and hence
\begin{equation}\label{eq:boundary-variation}
\frac{\delta S_R}{\delta q(\sigma)}
=
\frac{\delta W(T_Rq,R)}{\delta q(\sigma)}
-
\frac{\delta W(T_{-R}q,-R)}{\delta q(\sigma)}.
\end{equation}
In the local finite-order case these endpoint variations vanish for compactly supported variations away from the endpoints. In the nonlocal case they need not vanish, because \(W(T_{\pm R}q,\pm R)\) may still depend on \(q(\sigma)\).

The corresponding boundary mechanism was derived in \cite{HerediaThesis2023}. Related field-theoretic versions were formulated in \cite{Heredia2022, Heredia_2024_U1}. We restate the result in the present notation because it provides the precise link between nonlocal exactness and variational triviality.

\begin{theorem}[Boundary criterion]
\label{thm:boundary-criterion}
Let $\Lag(T_tq,t)=\D W(T_tq,t)$ be a nonlocal total derivative. Assume that the limit of the functional derivatives of the truncated actions exists and represents the first variation, namely,
\[
\Psi(q,\sigma)=
\lim_{R\to\infty}
\frac{\delta S_R}{\delta q(\sigma)}.
\]
Then
\begin{equation}\label{eq:boundary-criterion}
\Psi(q,\sigma)
=
\lim_{R\to\infty}
\left[
\frac{\delta W(T_Rq,R)}{\delta q(\sigma)}
-
\frac{\delta W(T_{-R}q,-R)}{\delta q(\sigma)}
\right].
\end{equation}
Consequently,
\[
\Psi(q,\sigma)\equiv0
\quad\Longleftrightarrow\quad
\lim_{R\to\infty}
\left[
\frac{\delta W(T_Rq,R)}{\delta q(\sigma)}
-
\frac{\delta W(T_{-R}q,-R)}{\delta q(\sigma)}
\right]
\equiv0 .
\]
\end{theorem}

\begin{proof}
Equation \eqref{eq:boundary-criterion} follows by taking the limit \(R\to\infty\) in \eqref{eq:boundary-variation}. The equivalence is immediate.
\end{proof}

The stronger condition
\begin{equation}\label{eq:sufficient-boundary}
\lim_{R\to\infty}
\frac{\delta W(T_Rq,R)}{\delta q(\sigma)}
=
\lim_{R\to\infty}
\frac{\delta W(T_{-R}q,-R)}{\delta q(\sigma)}
=0
\end{equation}
is sufficient, but not necessary \cite{Heredia_2024_U1, Heredia2022, HerediaThesis2023}. The two endpoint variations may fail to vanish separately and nevertheless cancel in the difference above. Thus, unlike in the formal infinite-order setting, the condition \(\Lag=\D W\) is not by itself a variational triviality criterion for nonlocal functionals: one must also control the corresponding boundary contribution.

\subsection{A nonlocal total derivative with nonzero EL}\label{sec:counterexample}

We revisit the convolutional model introduced in \cite{HerediaThesis2023}, where the Lagrangian, a nonlocal primitive, and the corresponding Euler-Lagrange expression were derived. Here we use this model to compare the nonlocal boundary obstruction with the formal infinite-order criterion of Section~\ref{sec:formal}, and then make this construction concrete by choosing a shifted Gaussian kernel with an explicit smooth trajectory for which the Euler-Lagrange expression does not vanish. Appendix~\ref{app:formal-convolution} analyses the same model from the formal infinite-derivative viewpoint.

Let $G\in\mathcal S(\mathbb R)$ be a real-valued Schwartz kernel \cite{Vladimirov}. For this convolutional model, we consider the translation-invariant kinematic space\footnote{The subscripts $b$ and $c$ stand for bounded and compact support, respectively.}
\[
\mathcal K_G:=C_b^1(\mathbb R)
=
\left\{
q\in C^1(\mathbb R):
\|q\|_\infty+\|\dot q\|_\infty<\infty
\right\},
\]
where $\|f\|_\infty:=\sup_{t\in\mathbb R}|f(t)|$, with admissible variations $\delta q\in C_c^\infty(\mathbb R)$. Since $\mathcal S(\mathbb R)\subset L^1(\mathbb R)$ and $q\in\mathcal K_G$ is bounded, the convolution
\[
(G*q)(t)
:=
\int_{\mathbb R}\dd\tau\,G(t-\tau)q(\tau)
\]
is absolutely convergent for every $t\in\mathbb R$. Indeed,
\[
\int_{\mathbb R}\dd\tau\,
|G(t-\tau)q(\tau)|
\leq
\|G\|_{L^1}\|q\|_\infty
<\infty,
\]
where $\|G\|_{L^1}:=\int_{\mathbb R}\dd s\,|G(s)|$. We then define
\begin{equation}\label{eq:convolution-L}
\Lag_G(T_tq):=\dot q(t)(G\conv q)(t),
\qquad
(G\conv q)(t):=\int_\R\dd\tau\,G(t-\tau)q(\tau).
\end{equation}
For $q\in\mathcal K_G$ and $\delta q\in C_c^\infty(\mathbb R)$, the first variation of the truncated action is
\[
\delta S_R[q;\delta q]
=
\int_{-R}^{R}\dd t\,
\left[
\delta\dot q(t)(G*q)(t)
+
\dot q(t)(G*\delta q)(t)
\right].
\]
Since $G\in\mathcal S(\mathbb R)\subset L^1(\mathbb R)$, $q$ and $\dot q$ are bounded by the definition of $\mathcal K_G$, and both $\delta q$ and $\delta\dot q$ belong to $L^1(\mathbb R)$ because they are smooth and compactly supported, the two terms in the first variation satisfy
\[
\begin{aligned}
&\int_{\mathbb R}\dd t\,
\left|\delta\dot q(t)(G*q)(t)\right|
+
\int_{\mathbb R}\dd t\,
\left|\dot q(t)(G*\delta q)(t)\right|
\\
&\qquad\leq
\|G\|_{L^1}
\left(
\|q\|_\infty\|\delta\dot q\|_{L^1}
+
\|\dot q\|_\infty\|\delta q\|_{L^1}
\right)
<\infty.
\end{aligned}
\]
Consequently, the limit of the first variation exists for every admissible compactly supported variation.

\begin{proposition}\label{prop:nonlocal-primitive}
The Lagrangian \eqref{eq:convolution-L} is a nonlocal total derivative. One primitive is
\begin{equation}\label{eq:nonlocal-W}
W_G(T_tq,t)=q(t)(G\conv q)(t)-\int_{\R^2}\dd\rho\,\dd\sigma\,[\theta(\rho)-\theta(\rho-t)]\dot G(\rho-\sigma)q(\rho)q(\sigma),
\end{equation}
where $\theta$ is the Heaviside function.
\end{proposition}

\begin{proof}
Since $G\in\mathcal S(\mathbb R)$ and $q\in\mathcal K_G$, the integral defining $W_G$ is absolutely convergent. Indeed,
\[
\begin{aligned}
&\int_{\mathbb R^2}\dd\rho\,\dd\sigma\,
|\theta(\rho)-\theta(\rho-t)|
|\dot G(\rho-\sigma)q(\rho)q(\sigma)|\leq
|t|\|\dot G\|_{L^1}\|q\|_\infty^2
<\infty.
\end{aligned}
\]
Furthermore, the integral term in \eqref{eq:nonlocal-W} can be written as $\int_0^t\dd\rho\,
q(\rho)(\dot G\conv q)(\rho)$. Differentiating the first term gives
\[
\frac{\dd}{\dd t}
\{q(t)(G\conv q)(t)\}
=
\dot q(t)(G\conv q)(t)
+
q(t)(\dot G\conv q)(t).
\]
By the fundamental theorem of calculus,
the derivative of the integral term is
$q(t)(\dot G\conv q)(t)$.
Therefore,
\[
\D W_G(T_tq,t)
=
\dot q(t)(G\conv q)(t)
=
\Lag_G(T_tq),
\]
as required.
\end{proof}

\begin{proposition}\label{prop:EL-convolution}
The nonlocal EL expression of \eqref{eq:convolution-L} is
\begin{equation}\label{eq:Psi-G}
\Psi_G(q,\sigma)=\int_\R\dd t\,\dot q(t)\,[G(t-\sigma)-G(\sigma-t)].
\end{equation}
In particular, $\Psi_G$ vanishes identically on $\mathcal K_G$ if and only if $G$ is even.
\end{proposition}

\begin{proof}
The functional derivative is\footnote{Functional derivatives are understood in the distributional sense \cite{Vladimirov}.}
\[
\frac{\delta \Lag_G(T_tq)}{\delta q(\sigma)}=\dot\delta(t-\sigma)(G\conv q)(t)+\dot q(t)G(t-\sigma).
\]
Integrating over $t$ gives
\[
\Psi_G(q,\sigma)=-(\dot G\conv q)(\sigma)+\int_\R\dd t\,\dot q(t)G(t-\sigma).
\]
Since $q$ is bounded and $G$ is a Schwartz function, integration by parts gives $-(\dot G*q)(\sigma) =
-\int_{\mathbb R}\dd t\, \dot q(t)G(\sigma-t)$. Substituting this identity into the previous expression yields \eqref{eq:Psi-G}. If $G$ is even, the bracket in \eqref{eq:Psi-G} vanishes identically.

Conversely, suppose that $G$ is not even and define $H(t):=G(t)-G(-t)$. Then $H$ is a nonzero real-valued Schwartz function. Choosing the admissible trajectory $q(t):=\dot H(t) \in\mathcal S(\mathbb R)\subset\mathcal K_G$, and evaluating \eqref{eq:Psi-G} at $\sigma=0$, we obtain
\[
\begin{aligned}
\Psi_G(q,0)
&=
\int_{\mathbb R}\dd t\,
\ddot H(t)H(t)=
-\int_{\mathbb R}\dd t\,
|\dot H(t)|^2
<0,
\end{aligned}
\]
where integration by parts is justified by the Schwartz decay of $H$. Thus, $\Psi_G$ is not identically zero
whenever $G$ is not even, proving the claimed equivalence.
\end{proof}

It is worth emphasizing that $\Lag_G(T_tq)$ has no explicit time dependence, whereas the nonlocal primitive $W_G(T_tq,t)$ given in \eqref{eq:nonlocal-W} does. However, this explicit dependence does not, by itself, determine variational triviality. Indeed, Appendix~\ref{app:boundary-check} verifies Theorem~\ref{thm:boundary-criterion} directly from \eqref{eq:nonlocal-W}, showing that the limit of the difference between the endpoint variations coincides with the nonlocal Euler-Lagrange expression \eqref{eq:Psi-G}. Thus, the obstruction to variational triviality is precisely the nonvanishing variational boundary contribution.

\begin{example}[Shifted Gaussian kernel]\label{ex:gaussian}
Choose
\begin{equation}\label{eq:shifted-gaussian}
G_{\mu,a}(s):=
\exp\left[-\frac{(s-\mu)^2}{2a^2}\right],
\qquad a>0,\qquad \mu\neq0.
\end{equation}
This kernel belongs to $\mathcal S(\mathbb R)$ and is not even for $\mu\neq0$. Inspired by the regularized Newtonian potential appearing in nonlocal gravity (see for instance, \cite{Biswas2012}), we consider the following test trajectory:
\begin{equation}\label{eq:qell}
q_\ell(t):=
\begin{cases}
\displaystyle \frac{\erf(t/2\ell)}{t},
& t\neq0,\\[0.8em]
\displaystyle \frac{1}{\sqrt\pi\,\ell},
& t=0,
\end{cases}
\qquad \ell>0.
\end{equation}
To verify that this trajectory is admissible, notice that it admits the integral representation
\[
q_\ell(t)=
\frac{1}{\sqrt\pi\,\ell}
\int_0^1\dd u\,
e^{-u^2t^2/(4\ell^2)}.
\]
This shows that $q_\ell$ is smooth, even, and bounded. Moreover,
\begin{equation}\label{eq:qell-prime}
\dot q_\ell(t)=
-\frac{t}{2\sqrt\pi\,\ell^3}
\int_0^1\dd u\,
u^2e^{-u^2t^2/(4\ell^2)}.
\end{equation}
Since $x e^{-x^2}$ is bounded, $\dot q_\ell$ is also bounded. Therefore, $q_\ell\in C_b^1(\mathbb R)=\mathcal K_G$. Furthermore, $\dot q_\ell$ is odd and $\dot q_\ell(t)<0$ for every $t>0$. Evaluating \eqref{eq:Psi-G} at $\sigma=0$ gives 
\begin{equation}\label{eq:Psi-zero}
\Psi_{\mu,a}(q_\ell,0)
=
\int_\R\dd t\,\dot q_\ell(t)
[G_{\mu,a}(t)-G_{\mu,a}(-t)].
\end{equation}
The antisymmetric part of the shifted Gaussian can be written exactly as
\begin{equation}\label{eq:G-odd-difference}
G_{\mu,a}(t)-G_{\mu,a}(-t)
=
2e^{-(t^2+\mu^2)/(2a^2)}
\sinh\left(\frac{\mu t}{a^2}\right).
\end{equation}
Since both $\dot q_\ell(t)$ and the bracket in \eqref{eq:Psi-zero} are odd, their product is even. Consequently,
\[
\Psi_{\mu,a}(q_\ell,0)
=
4e^{-\mu^2/(2a^2)}
\int_0^\infty\dd t\,
\dot q_\ell(t)e^{-t^2/(2a^2)}
\sinh\left(\frac{\mu t}{a^2}\right).
\]
The integral is absolutely convergent because $\dot q_\ell$ is bounded and the difference of Gaussian kernels belongs to $\mathcal S(\mathbb R)$. For $t>0$, we have $\dot q_\ell(t)<0$, whereas
$\sinh(\mu t/a^2)$ has the sign of $\mu$.
Thus,
\[
\operatorname{sgn}
\Psi_{\mu,a}(q_\ell,0)
=
-\operatorname{sgn}(\mu),
\]
and hence
\begin{equation}\label{eq:nonzero-Psi}
\Psi_{\mu,a}(q_\ell,0)\neq0
\qquad
\text{for all }a,\ell>0,\quad\mu\neq0.
\end{equation}
Therefore, the Lagrangian \eqref{eq:convolution-L}, with the shifted Gaussian kernel, is a nonlocal total derivative whose nonlocal Euler-Lagrange expression does not vanish identically. See Appendix~\ref{app:formal-convolution} for the corresponding formal viewpoint.
\end{example}

\section{Discussion}\label{sec:discussion}

We have examined the status of total derivatives in two distinct variational frameworks: the formal jet algebra and the functional formulation of nonlocal mechanics. In the autonomous formal setting considered here, Theorem \ref{th:cfc} provides a characterization of variationally trivial Lagrangians:
\[
    E(L)\equiv 0
    \quad\Longleftrightarrow\quad
    L=c+DW,
    \qquad c\in\mathbb{R},\quad W\in\widehat{\mathcal A}.
\]
For each homogeneous component of polynomial degree $r\geq 1$, this condition is equivalent to divisibility of its symmetric symbol by $x_1+\cdots+x_r$. These equivalences hold coefficientwise in the formal algebra, independently of convergence of the infinite derivative series as a functional on trajectories.

The situation is different for nonlocal functionals. As established in the nonlocal Lagrangian formalism of \cite{Heredia_2024_U1, HerediaThesis2023}, the existence of a nonlocal primitive is not sufficient to guarantee variational triviality. Indeed, within the sufficiently regular class considered in Theorem \ref{thm:universal-nonlocal-primitive}, every nonlocal Lagrangian admits a representation of the form $L=DW$. The relevant criterion is instead the cancellation of the variational boundary contribution, as expressed in Theorem \ref{thm:boundary-criterion}. Under the stated assumptions on functional derivatives and limits, this cancellation is necessary and sufficient for the nonlocal Euler-Lagrange expression to vanish identically. The stronger requirement that both
endpoint variations vanish separately is sufficient, but not necessary.

The convolutional model illustrates this distinction explicitly. For a real Schwartz kernel $G$, Proposition \ref{prop:EL-convolution} shows that the nonlocal Euler-Lagrange expression vanishes identically on the chosen trajectory space if and only if $G$ is even. In particular, the shifted Gaussian provides a smooth nonlocal total derivative with a nonvanishing Euler-Lagrange expression for every nonzero shift. Appendix \ref{app:formal-convolution} shows, however, that this example is not a case in which formal variational triviality is lost through summation: for a nonzero shift, the corresponding formal Euler-Lagrange expression is already nonzero. 

These results emphasize that the implications of total-derivative exactness differ between the two variational frameworks. In the formal infinite-order setting, variational triviality is characterized by algebraic factorization, independently of convergence. For nonlocal functionals, however, the existence of a primitive is not sufficient: variational triviality requires the cancellation of the corresponding
variational boundary contribution. Consequently, discarding a nonlocal total derivative may change the equations of motion unless this boundary condition is verified within the chosen trajectory space.

\ack{The author thanks Josep Llosa for discussions on nonlocal variational principles, and his daughter Daniela for sleeping while he was writing the manuscript.}

\data{No data were generated or analysed in this work.}

\appendix

\section*{Appendix}

\section{Boundary check for the convolutional example}
\label{app:boundary-check}

We briefly verify Theorem~\ref{thm:boundary-criterion} for the convolutional example. Write $H_t(\rho):=\theta(\rho)-\theta(\rho-t)$, so that the primitive \eqref{eq:nonlocal-W} reads
\[
W_G(T_tq,t)
=
q(t)(G*q)(t)
-
\int_{\mathbb R^2}d\rho\,d\eta\,
H_t(\rho)\dot G(\rho-\eta)q(\rho)q(\eta).
\]
A direct functional differentiation gives
\[
\begin{aligned}
\frac{\delta W_G(T_tq,t)}{\delta q(\sigma)}
&=
\delta(t-\sigma)(G*q)(t)+q(t)G(t-\sigma)
\\
&\quad
-H_t(\sigma)(\dot G*q)(\sigma)
-\int_{\mathbb R}d\rho\,H_t(\rho)\dot G(\rho-\sigma)q(\rho).
\end{aligned}
\]
When paired with any compactly supported variation $\delta q$, the endpoint delta distributions contribute $\delta q(\pm R)$, which vanish for sufficiently large $R$. Moreover, since $q$ is bounded and $G\in\mathcal S(\mathbb R)$, the terms $q(\pm R)G(\pm R-\sigma)$ vanish as $R\to\infty$, uniformly for $\sigma$ in compact sets. Moreover,
\[
H_R=\mathbf 1_{(0,R)},\qquad
H_{-R}=-\mathbf 1_{(-R,0)},
\]
hence
\[
H_R-H_{-R}=\mathbf 1_{(-R,R)}\longrightarrow 1.
\]
Therefore
\[
\begin{aligned}
\lim_{R\to\infty}
\left[
\frac{\delta W_G(T_Rq,R)}{\delta q(\sigma)}
-
\frac{\delta W_G(T_{-R}q,-R)}{\delta q(\sigma)}
\right]
&=
-(\dot G*q)(\sigma)
-\int_{\mathbb R}d\rho\,\dot G(\rho-\sigma)q(\rho)
\\
&=
\int_{\mathbb R}d\rho\,\dot q(\rho)
\bigl[G(\rho-\sigma)-G(\sigma-\rho)\bigr].
\end{aligned}
\]
Thus the boundary difference is exactly
\[
\Psi_G(q,\sigma)
=
\int_{\mathbb R}d\rho\,\dot q(\rho)
\bigl[G(\rho-\sigma)-G(\sigma-\rho)\bigr]\,,
\]
as expected. Consequently, the obstruction is precisely the non-vanishing of the variational boundary contribution of the nonlocal primitive.

\section{The convolution example in the formal picture}
\label{app:formal-convolution}

Let \(\Lag_G(T_tq)\) be the convolutional Lagrangian defined in \eqref{eq:convolution-L}. A formal Taylor expansion gives
\[
(G*q)(t)\sim \sum_{n\geq0}a_n q_n(t),
\qquad
a_n:=\frac{(-1)^n}{n!}\int_{\mathbb R}ds\,s^nG(s),
\qquad q_n:=D^nq.
\]
It is important to emphasize that this expansion is purely formal. Although $G\in\mathcal S(\mathbb R)$ guarantees that all coefficients $a_n$ are well defined, it does not imply that the resulting infinite-derivative series represents the original convolution operator. Such an identification requires additional analytical justification. Appendix \ref{app:infser-conv} provides an explicit example of a nonzero Schwartz kernel for which all coefficients $a_n$ vanish, while the corresponding convolution operator remains nontrivial.

Thus,
\[
L_G^{\rm form}=q_1A(D)q,
\qquad
A(z):=\sum_{n\geq0}a_nz^n.
\]
In symmetric quadratic form its generative series is
\[
C(x,y)=xA(y)+yA(x).
\]
By Theorem \ref{th:qic}, \(L_G^{\rm form}\) is a formal total derivative if and only if $C(x,-x)=0$. Since
\[
C(x,-x)=x\bigl(A(-x)-A(x)\bigr),
\]
we obtain
\[
L_G^{\rm form}=DW_{\rm form}
\quad\Longleftrightarrow\quad
A(-x)=A(x).
\]
Equivalently,
\[
\EL(L_G^{\rm form})
=
D\bigl(A(-D)-A(D)\bigr)q.
\]

For the shifted Gaussian $G_{\mu,a}(s)$ defined in (\ref{eq:shifted-gaussian}), one finds
\[
A_{\mu,a}(z)
=
\int_{\mathbb R}ds\,G_{\mu,a}(s)e^{-zs}
=
\sqrt{2\pi}\,a\,
\exp\!\left(-\mu z+\frac{a^2z^2}{2}\right).
\]
Hence
\[
A_{\mu,a}(-z)-A_{\mu,a}(z)
=
2\sqrt{2\pi}\,a\,e^{a^2z^2/2}\sinh(\mu z),
\]
which is not identically zero when \(\mu\neq0\). Therefore
\[
\EL(L_{\mu,a}^{\rm form})\not\equiv0.
\]
For the shifted Gaussian kernel with $\mu\neq 0$, the nonlocal Lagrangian admits a primitive $W_G$, as shown in Section~\ref{sec:counterexample}. However, its formal
infinite-derivative expansion does not admit a primitive in the formal jet algebra of Section~\ref{sec:formal}, since $A(-z)\neq A(z)$. Consequently, $E(L_G^{\mathrm{form}})\not\equiv 0$. There is no contradiction: the existence of a nonlocal primitive does not imply formal exactness. 

\section{A Schwartz kernel with vanishing moments.}\label{app:infser-conv}

The formal derivative expansion does not necessarily determine the underlying convolution operator, even for Schwartz kernels. To illustrate this, let $\chi\in C_c^\infty((1,2))$ be a nonzero real-valued function, and define a kernel $G$ through its Fourier transform
\[
\widehat G(\omega)
=
i\bigl[\chi(\omega)-\chi(-\omega)\bigr],
\]
where we use the convention
\[
\widehat G(\omega)
=
\int_{\mathbb R}\dd t\,e^{-i\omega t}G(t).
\]
Since $\widehat G\in C_c^\infty(\mathbb R)\subset\mathcal S(\mathbb R)$, and the Fourier
transform is an isomorphism of $\mathcal S(\mathbb R)$ onto itself \cite{Vladimirov}, it follows that $G=\mathcal F^{-1}\widehat G\in\mathcal S(\mathbb R)$. Moreover, the Fourier transform is purely imaginary and odd, so $G$ is real, odd, and nonzero\footnote{The inverse of the Fourier transform is: $G(t) = - \frac{1}{\pi}\int^2_1 d\omega\;\chi(\omega)\sin(\omega t)$.}.

By construction, $\widehat G$ vanishes in a neighbourhood of $\omega=0$. Consequently, all moments of $G$ vanish:
\[
\int_{\mathbb R}\dd s\,s^nG(s)
=
i^n\widehat G^{(n)}(0)
=0,
\qquad n\geq0.
\]
It follows that every coefficient in the formal derivative expansion introduced above vanishes:
\[
a_n
=
\frac{(-1)^n}{n!}
\int_{\mathbb R}\dd s\,s^nG(s)
=0,
\qquad n\geq0.
\]
Hence, as formal power series,
\[
A(z)\equiv0,
\qquad
L_G^{\mathrm{form}}=q_1A(D)q\equiv0,
\qquad
E(L_G^{\mathrm{form}})\equiv0.
\]
Nevertheless, the nonlocal Euler-Lagrange expression need not vanish. Indeed, consider the admissible trajectory
\[
q(t)=G'(t)
\in\mathcal S(\mathbb R)
\subset\mathcal K_G.
\]
Using the nonlocal Euler-Lagrange expression derived in Proposition~\ref{prop:EL-convolution}, evaluated at $\sigma=0$, we obtain
\[
\begin{aligned}
\Psi_G(q,0)
&=
\int_{\mathbb R}\dd t\,G''(t)
\bigl[G(t)-G(-t)\bigr]
\\
&=
2\int_{\mathbb R}\dd t\,G''(t)G(t)
\\
&=
-2\int_{\mathbb R}\dd t\,|G'(t)|^2
<0,
\end{aligned}
\]
where we have used the oddness of $G$ and integration by parts, with vanishing boundary terms due to Schwartz decay. By continuity, \(\Psi_G(q,\sigma)\) has the same sign as \(\Psi_G(q,0)\) for \(\sigma\) sufficiently close to \(0\). We can therefore choose a nonzero, nonnegative, smooth variation \(\delta q\), compactly supported in this region, such that
\[
\delta S[q;\delta q]
=
\int_{\mathbb R}\dd\sigma\,
\Psi_G(q,\sigma)\delta q(\sigma)
<0.
\]

Thus, the moment-based formal Euler-Lagrange expression vanishes identically, whereas the Euler-Lagrange expression of the corresponding convolutional Lagrangian does not.

This does not represent a failure of formal variational triviality. Rather, it shows that the moment-based formal expansion need not represent the original convolution operator on the chosen trajectory space. In particular, the vanishing formal series cannot be identified with the nonzero integral operator without additional justification of the
interchange between the Taylor expansion and the convolution integral. This distinction is consistent with the domain and convergence issues discussed in \cite{HerediaLlosa2024IDIO}.

\printbibliography

\end{document}